\documentclass[aps,prd,amsmath,amsfonts,a4paper,reprint,preprintnumbers,twocolumn,nofootinbib,superscriptaddress,10pt]{revtex4-1}

\usepackage{natbib}
\usepackage{amsmath}
\usepackage{amsfonts}
\usepackage{amssymb}
\usepackage{color}
\usepackage{bm}

\usepackage{graphicx}

\usepackage[labelfont={bf,scriptsize},textfont={scriptsize},justification=RaggedRight,format=hang]{caption,subfig}

\usepackage[T1]{fontenc}

\newcommand{\mpl}{M_{\mathrm{P}}}

\def\stu{St\"uckelberg}

\newcommand{\K}{\mathcal{K}}

\newcommand{\G}{\mathcal{G}}

\newcommand{\T}{\mathcal{T}}

\newcommand{\chris}[2]{\mathit{\Gamma}\!\!_{#1}\!{}^{#2}}

\newcommand{\Tr}[1]{[{#1}]}

\def\<{\left<}
\def\>{\right>}
\def\({\left(}
\def\){\right)}

\usepackage{mathtools}
\usepackage{ulem}

\def \bal#1\eal  {\begin{align} #1 \end{align}}
\newcommand{\be} {\begin{equation}}
\newcommand{\ee} {\end{equation}}
\newcommand{\nn} {\nonumber\\}

\newcommand{\pd} {\partial}

\newcommand{\mc} {\mathcal}

\newcommand{\ai}{{\alpha}}
\newcommand{\bi}{{\beta}}

\newcommand{\ri}{{\rho}}
\newcommand{\si}{{\sigma}}

\definecolor{dullpurple}{rgb}{0.431,0.188,0.534}
\definecolor{darkgreen}{rgb}{0.133,0.545,0.133}

\RequirePackage[colorlinks=true
,urlcolor=dullpurple
,anchorcolor=blue
,citecolor=dullpurple
,filecolor=blue
,linkcolor=darkgreen
,menucolor=blue
,pagecolor=blue
,linktocpage=true
,pdfproducer=medialab
]{hyperref}

\begin{document}

\title{On the uniqueness of the non-minimal matter coupling in massive gravity and bigravity}

   \author{Qing-Guo Huang}
   \email{huangqg@itp.ac.cn}
   \affiliation{State Key Laboratory of Theoretical Physics, 
	Institute of Theoretical Physics, \\
	Chinese Academy of Sciences, P.O. Box 2735, Beijing 100190, China}
   
    \author{Raquel H. Ribeiro}
	\email{R.Ribeiro@qmul.ac.uk}
	\affiliation{School of Physics and Astronomy, Queen Mary University of London, \\
Mile End Road, London, E1 4NS, U.K.}
	\affiliation{CERCA/Department of Physics, Case Western Reserve University,
	\\ 10900 Euclid Ave, Cleveland, OH 44106, U.S.A.}

	\author{Yu-Hang Xing}
	\email{xingyh@itp.ac.cn}
	\affiliation{State Key Laboratory of Theoretical Physics, 
	Institute of Theoretical Physics, \\
	Chinese Academy of Sciences, P.O. Box 2735, Beijing 100190, China}

		 \author{Ke-Chao Zhang}
	\email{zkc@itp.ac.cn}
	\affiliation{State Key Laboratory of Theoretical Physics, 
	Institute of Theoretical Physics, \\
	Chinese Academy of Sciences, P.O. Box 2735, Beijing 100190, China}

 \author{Shuang-Yong Zhou}
	\email{sxz353@case.edu}
	\affiliation{CERCA/Department of Physics, Case Western Reserve University,
	\\ 10900 Euclid Ave, Cleveland, OH 44106, U.S.A.}

\begin{abstract}

In de Rham--Gabadadze--Tolley (dRGT) massive gravity and bi-gravity, a non-minimal matter coupling involving both metrics generically re-introduces the Boulware--Deser (BD) ghost. A non-minimal matter coupling via a simple, yet specific composite metric has been proposed, which eliminates the BD ghost below the strong coupling scale.  Working explicitly in the metric formulation and for arbitrary spacetime dimensions, we show that this composite metric is the unique consistent non-minimal matter coupling below the strong coupling scale, which emerges out of two diagnostics, namely, absence of Ostrogradski ghosts in the decoupling limit and absence of the BD ghost from matter quantum loop corrections. 

\end{abstract}
	
\maketitle

	\section{Introduction}
	\label{sec:introduction}
	Developing consistent theories of gravity where the graviton has a mass
	has seen renewed interest especially in the last few years~\cite{Hinterbichler:2011tt,deRham:2014zqa}. Many studies have been devoted to constructing a non-linear extension to the
	Fierz--Pauli mass term~\cite{Fierz:1939ix} 
	while keeping the theory manifestly ghost-free at 
	the classical level.
	The non-linear completion was accomplished only a few years ago for massive 
	gravity~\cite{deRham:2010ik,deRham:2010kj,Hassan:2011ea,Hassan:2011hr} 
	and then for 
	bigravity~\cite{Hassan:2011zd} (where both metrics have their own kinetic term). Consequently, 
	an array of phenomenological explorations followed which have focused on studying 
	cosmological applications of these theories. 
	
	Recently, other questions
	in the context of these Effective Field Theory	formulations of massive gravity and bigravity have been raised.
	Since these theories are highly non-linear and diffeomorphism invariance is explicitly broken, 
	it might not be immediately apparent whether the coefficients
	that determine the interactions are natural in the t'Hooft sense. 
	In fact, these couplings are protected by a quantum analogue of screening
	sourced by the Vainhstein 
	effect~\cite{deRham:2012ew,deRham:2013qqa}. There are reasons to believe this is, in fact, a feature of all theories which exhibit a Vainshtein mechanism~\cite{Vainshtein:1971ip}---see, for example, Refs.~\cite{Nicolis:2004qq,deRham:2012ew,Brouzakis:2013lla,Brouzakis:2014bwa,deRham:2014wfa}.
	
	 The nature of the interactions
	originated at the quantum level has motivated research into 
	developing consistent couplings to the matter sector~\cite{Hassan:2011zd,Hassan:2012wr,Akrami:2013ffa,Schmidt-May:2014xla,Deser:2013rxa,Heisenberg:2014rka}
	which would not excite the Boulware--Deser (BD) ghost~\cite{deRham:2014naa,deRham:2014fha,Hinterbichler:2015yaa}.
	Since the phenomenology of these theories is 
	explicitly dependent on the way the massive graviton
	couples to matter, it is important to detail which couplings are 
	theoretically consistent, both classically and quantum mechanically, before propagating them towards observational
	tests.

While most works have investigated cosmological solutions with a minimal coupling to one of the metrics, it is possible that a matter coupling built out of an admixture of both metrics admits a richer cosmological phenomenology.
Unfortunately, constructing such a coupling to a matter field generically re-introduces the BD ghost~\cite{Yamashita:2014fga,deRham:2014naa,Soloviev:2014eea} already at the classical level. Nevertheless, an exception has been pointed out in the literature: by combining the two metrics appropriately, one can eliminate the BD ghost below the strong coupling scale \cite{deRham:2014naa,deRham:2014fha}\footnote{In 4 dimensions, the strong coupling scale is the usual $\Lambda_3\equiv (m^2\mpl)^{1/3}$, while in $D$ dimensions it becomes $\Lambda_{D-1}\equiv(m^2\mpl^{(D-2)/2})^{2/(D+2)} $.}---see also Ref.~\cite{Noller:2014sta} for a complementary derivation in the vielbein language. Ideally, one would want a matter coupling where the BD ghost is eliminated fully, in which the strong coupling scale is not necessarily a physical cut-off, and thus above the strong coupling scale the interactions would be redressed owing to the Vainshtein mechanism. It has been argued that this may be achieved by switching to the vielbein formulation and relaxing the symmetric vielbein condition~\cite{Hinterbichler:2015yaa}. However, additional considerations have revealed that this is not possible for the effective composite metric of Ref.~\cite{deRham:2015cha}. 

Therefore, given the recent and active investment in exploring the ghost-freeness of this matter coupling~\cite{deRham:2014naa, Noller:2014sta, Hassan:2014gta,deRham:2014fha,Soloviev:2014eea,Hinterbichler:2015yaa,deRham:2015cha},  
it is important to understand whether there are other non-minimal matter couplings which are also ghost-free below or even above the strong coupling scale. In this paper, we prove that the non-minimal coupling originally proposed in Ref.~\cite{deRham:2014naa} is the \textit{unique} composite metric that avoids the BD ghost below the strong coupling scale. This has profound consequences on the phenomenology of these theories~\cite{Gumrukcuoglu:2014xba,Mukohyama:2014rca,Comelli:2015pua,Enander:2014xga} since it limits the choice of matter couplings in these theories to at most two free parameters for each matter sector.

\section{Uniqueness of the composite metric}
\label{sec:unicm}

We will be working in $D$ dimensions and consider a matter coupling of the form
\be
\label{nmc}
\mc{L}_{matter} = \sum_{I} \mc{L}_m(G^{(I)}_{\mu\nu},\psi^{(I)},\pd \psi^{(I)})   ,
\ee
where $G^{(I)}_{\mu\nu}$ are effective, composite metrics for the $I$-th matter sector denoted collectively as $\psi^{(I)}$. One or both of the two metrics $g_{\mu\nu}$ and $f_{\mu\nu}$ are assumed to have a standard kinetic term, corresponding to massive gravity or bigravity respectively. We shall consider a strictly local composite metric, by which we mean that we only consider point-wise operations, including inverting $g_{\mu\nu}$ and $f_{\mu\nu}$, in constructing $G^{(I)}_{\mu\nu}(x)$ out of $g_{\mu\nu}(x)$ and $f_{\mu\nu}(x)$, but not their derivatives or non-local operations. The dRGT graviton potential is given by
\be
\label{dRGTm}
\mc{U} = \sum_{s=0}^D \alpha_s U_s(\mc{K}), ~~ {\rm with}~ U_s(\mc{K}) = \mc{K}^{\mu_1}_{[\mu_1} \mc{K}^{\mu_2}_{\mu_2} \cdots \mc{K}^{\mu_s}_{\mu_s]}   ,
\ee
where $\mc{K}^\mu_\nu$ is defined by
\be
f_{\mu\nu} = g_{\mu\ri} (\delta^\ri_{\si}-\mc{K}^\ri_{\si})(\delta^\si_{\nu}-\mc{K}^\si_{\nu})  ,
\ee 
with the branch choice such that $\mc{K}^\mu_\nu\to 0$ when $g_{\mu\nu}\to f_{\mu\nu}$.
$\mc{K}^\mu_\nu$ can be viewed as the deviation of $f_{\mu\nu}$ from $g_{\mu\nu}$. Without loss of generality, we can choose $g_{\mu\nu}$ and $\mc{K}^\mu_\nu$ as the elementary building blocks to construct $G^{(I)}_{\mu\nu}$. Then the most general composite metric is given by
\be
G^{(I)}_{\mu\nu}  = \sum_N G^{(I)}{}^{N}_{\mu\nu}   ,
\ee
with
\begin{equation}
\label{eq:generalansatz}
G^{(I)}{}^{N}_{\mu\nu}= g_{\mu\ri} \sum_{n=0}^N p^{(I)}_{N,n}([\mc{K}],[\mc{K}^2],...)(\K^{n})^{\ri}_{\nu}  \ ,
\end{equation}
where $[\,\,]$ is the trace of the matrix enclosed, $p^{(I)}_{N,n}([\mc{K}],[\mc{K}^2],...)$ are arbitrary functions of the various traces of matrix $\mc{K}^\mu_\nu$,  $(\mc{K}^2)^\ri_\nu \equiv \mc{K}^\ri_\si \mc{K}^\si_\nu$ and so on.  Note that $p^{(I)}_{(N,n)}$ is $(N-n)$-th order in $\mc{K}^\mu_\nu$. Since $\mc{K}_{\mu\nu}\equiv  g_{\mu\ri}\mc{K}^{\ri}_{\nu}$ is symmetric in its indices, the indices $\mu$ and $\nu$ in Eq.~(\ref{eq:generalansatz}) are symmetrized implicitly.

The strategy of our proof is to impose two consistency conditions at different steps to restrict the form of $G^{(I)}_{\mu\nu}$ to that of Ref.~\cite{deRham:2014naa}:
\be
G^{(I)}_{\mu\nu} = g_{\mu\ri} \big( \alpha_{(I)}^2 \delta^\ri_\nu+2 \alpha_{(I)}\beta_{(I)}\mc{K}^\ri_\nu + \beta_{(I)}^2(\mc{K}^2)^\ri_\nu \big)   ,
\label{eq:finalansatz}
\ee  
where $\alpha_{(I)}$ and $\beta_{(I)}$ are constant. When $\beta_{(I)}=0$ or $\beta_{(I)}=-\alpha_{(I)}$, a minimal matter coupling is reproduced.

\subsection{No Ostrogradski ghosts in the decoupling limit}

We will first impose the condition that the non-minimal matter coupling (\ref{nmc}) does not give rise to Ostrogradski instabilities~\cite{Ostrogradsky} for the scalar mode in the decoupling limit. The precise meaning of this limit can be found, for example, in Ref.~\cite{deRham:2014zqa}. But for our purposes this means that we focus on the scalar \stu \ mode, taking the limit
\bal
g_{\mu\nu} &\to \eta_{\mu\nu}     ,
\\
f_{\mu\nu} & \to \pd_\mu\phi^a\pd_\nu\phi^b\eta_{ab}, ~~~{\rm with} ~~~\phi^a = x^a -\eta^{a\mu}\pd_\mu \pi\ ,
\eal
where $\pi$ is the scalar \stu \ mode. In this limit, we have 
\be
\mc{K}^\mu_\nu = \Pi^\mu_\nu = \pd^\mu\pd_\nu \pi   ,
\ee
whilst the tensor and vector modes are suppressed. We will use  $\eta_{\mu\nu}$ to lower the indices in this subsection. Consequently, we will use $\mc{K}^\mu_\nu$ and $\Pi^\mu_\nu$ interchangeably in this subsection. $G^{(I)}_{\mu\nu}$ are now simply functions of $\Pi_{\mu\nu}$ (and $\eta_{\mu\nu}$). 

To avoid the Ostrogradski ghost~\cite{Ostrogradsky}, we shall require the contribution to the $\pi$ equation of motion coming from $\mc{L}_{matter}$ to not contain higher order derivatives, either in $\pi$ or in the matter fields. Since the contributions from different $\mc{L}_m(G^{(I)}_{\mu\nu},\psi^{(I)},\pd \psi^{(I)})$ contain different matter fields, these different contributions do not cancel each other in the $\pi$ equation of motion.  This implies that we can focus on one matter sector, and, omit the index $I$ here and afterwards. Thus, the contributions to the $\pi$ equation of motion arising from $\mc{L}_m(G_{\mu\nu},\psi,\pd \psi)$,
\begin{align}
\label{eq:eompiRefined}
\mc{E}_\pi&=\pd_\rho\pd_\sigma\left[\sqrt{-G}\,\,T^{\mu\nu}\frac{\pd G_{\mu\nu}}{\pd\Pi_{\rho\sigma}}\right]	\nn
&=\pd_\rho\pd_\sigma(\sqrt{-G}\,\,T^{\mu\nu})\frac{\pd G_{\mu\nu}}{\pd\Pi_{\rho\sigma}} + 2\pd_{(\rho}(\sqrt{-G}\,\,T^{\mu\nu})\pd_{\sigma)}\frac{\pd G_{\mu\nu}}{\pd\Pi_{\rho\sigma}}\nn
&~~~+(\sqrt{-G}\,\,T^{\mu\nu})\, \pd_\rho\pd_\sigma\frac{\pd G_{\mu\nu}}{\pd\Pi_{\rho\sigma}}   ,
\end{align}
should not contain higher order derivatives, where $G$ is the determinant of $G_{\mu\nu}$ and the energy momentum tensor from the $I$-th matter sector is given by
\be
T^{\mu\nu}=\frac{-2}{\sqrt{-G}}\frac{\pd \mc{L}_m(G_{\mu\nu},\psi,\pd \psi)}{\pd G_{\mu\nu}}\label{eq:eompi}.
\ee 
It is usually assumed that the matter sectors are diffeomorphism invariant separately, so we have the energy momentum conservation for each sector 
\begin{equation}
\label{consequ}
\pd_\nu(\sqrt{-G}T^{\mu\nu})+\chris{\nu\rho}{\mu}\sqrt{-G}\,T^{\nu\rho}=0\ ,
\end{equation}
where $\chris{\nu\rho}{\mu}$ is the Christoffel coefficients associated with the metric $G_{\mu\nu}$. Now, since $T^{\mu\nu}$ contains terms with first (or higher) derivatives of the matter fields, the first term of Eq. (10) contains terms with third (or higher) derivatives of the matter fields, which cannot be canceled by other terms. Therefore, as a necessary condition, we impose the first term of Eq.~\eqref{eq:eompiRefined} to vanish identically:
\be
\label{firstTzero}
\pd_\rho\pd_\sigma(\sqrt{-G}\,\,T^{\mu\nu})\frac{\pd G_{\mu\nu}}{\pd\Pi_{\rho\sigma}} =0    .
\ee 
For later convenience, we define
\begin{align}
\T^{\mu\nu}{}_{\rho\sigma} &= \pd_\rho\pd_\sigma(\sqrt{-G}\,\,T^{\mu\nu})\ ,	\\
\T^{\mu\nu}{}_{\rho} &= \pd_\rho(\sqrt{-G}\,\,T^{\mu\nu})	\ , \\
\T^{\mu\nu} &= \sqrt{-G}\,\,T^{\mu\nu}	\ , \ \textrm{and}\\
\G_{\mu\nu}{}^{\rho\sigma} &= \frac{\pd G_{\mu\nu}}{\pd \Pi_{\rho\sigma}}\ .
\end{align}
Since $\mc{T}^{\mu\nu}{}_{\rho}$ and $\mc{T}^{\mu\nu}{}_{\rho\sigma}$ are first and second derivatives of $\mc{T}_{\mu\nu}$ respectively, their numerical values at a specific, arbitrarily chosen point in spacetime would be independent from that of $\mc{T}_{\mu\nu}$ in the absence of Eq.~\eqref{consequ}. That is, by choosing the matter configuration appropriately in the neighborhood of a specific point, the numerical values of $\mc{T}^{\mu\nu}{}_{\rho\sigma}, \mc{T}^{\mu\nu}{}_{\rho}$ and $\mc{T}^{\mu\nu}$ can be assigned independently at that point, subject to the following constraints
\begin{subequations}
\begin{align}
\mc{T}^{\mu\nu}{}_{\rho\nu} &= -\mc{T}^{\sigma\nu}{}_\rho \chris{\sigma\nu}{\mu} - \mc{T}^{\sigma\nu}\pd_\rho\chris{\sigma\nu}{\mu}   ,
	\\
\mc{T}^{\mu\nu}{}_\nu &= -\mc{T}^{\nu\rho}\chris{\nu\rho}{\mu}   .
\end{align}
\end{subequations}

To simplify our discussion, we choose a matter configuration where $\mc{T}^{\mu\nu}{}_{\rho}$ and $\mc{T}^{\mu\nu}$ vanish at a spacetime point. This can always be achieved as follows: Suppose there are two matter configurations where $\mc{T}'^{\mu\nu}{}_{\rho}|_p=\mc{T}^{\mu\nu}{}_{\rho}|_p$ and $\mc{T}'^{\mu\nu}|_p=\mc{T}^{\mu\nu}|_p$, where $p$ is a spacetime point; That is,  $\Delta \mc{T}^{\mu\nu}{}_{\rho}|_p=0$ and $\Delta \mc{T}^{\mu\nu}|_p=0$; Then, one takes $\Delta \mc{E}_\pi=-2\pd_\rho\pd_\sigma\,[\, \Delta \mc{T}^{\mu\nu} \pd G_{\mu\nu}/\pd\Pi_{\rho\sigma}]$ as our starting $\mc{E}_\pi$. After this, at that point the constraint system reduces to a condition that is much easier to handle:
\be
\label{constT4}
\mc{T}^{\mu\nu}{}_{\rho\nu}=0   .
\ee 

To extract the conditions on $G_{\mu\nu}$ encoded in Eq.~\eqref{firstTzero}, we need to project out the traces of $\mc{T}^{\mu\nu}{}_{\rho\sigma}$. That is, we need a projector $\mc{P}$ such that $(\mc{P}\mc{T})^{\mu\nu}{}_{\rho\nu}$ vanishes for unconstrained $\mc{T}^{\mu\nu}{}_{\ri\si}$. This reduces the system of Eqs.~\eqref{firstTzero} and \eqref{constT4} to a single equation:
\be
\label{GPT}
\G_{\mu\nu}{}^{\rho\sigma} (\mc{P}\mc{T})^{\mu\nu}{}_{\rho\sigma} =0   .
\ee
One can shift the projector $\mc{P}$ to act on ${\mc{G}}_{\mu\nu}{}^{\rho\sigma}$ instead and leave $\mc{T}^{\mu\nu}{}_{\rho\sigma}$ to be a generic tensor. Then, getting rid of the generic tensor $\mc{T}^{\mu\nu}{}_{\rho\sigma}$, Eq.~\eqref{GPT} reduces to the requirement:
\be
\label{idG}
\widetilde{\mc{G}}_{\mu\nu}{}^{\rho\sigma} =(\mc{P}\mc{G})_{\mu\nu}{}^{\rho\sigma}= 0    .
\ee 
In $D$ dimensions, such a projector is explicitly given by\footnote{See Appendix \ref{app:projop} for details on the derivation.}
\begin{align}\label{eq:projection}
\widetilde{\mc{G}}_{\mu\nu}{}^{\rho\sigma}&=\mc{G}_{\mu\nu}{}^{\rho\sigma}-\frac{4}{D+2}\delta_{(\mu}^{(\rho}\mc{G}_{\nu)\gamma}{}^{\sigma)\gamma}\nn
&~~~~~+\frac{2}{(D+2)(D+1)}\delta_{(\mu}^{\rho}\delta_{\nu)}^{\sigma}\mc{G}_{\ai\bi}{}^{\ai\bi}    .
\end{align}
In other words, Eq.~\eqref{idG} has to be an identity. We will make use of this identity to constrain the form of $G_{\mu\nu}$. 

Now, since this is an identity, different orders of $\Pi$, and thus different orders of $\K$, should cancel separately, so it is sufficient to consider the $N$-th order terms of the general ansatz
\begin{equation}\label{eq:NndOrderG}
G^{N}_{\mu\nu}=g_{\mu\ri}\sum_{n=0}^N p_{N,n}([\K],[\K^2],...)(\K^{n})^{\ri}_{\nu}    ,
\end{equation}
where $p_{N,n}([\K],[\K^2],...)$ are of order $\mc{O}(\mc{K}^{N-n})$.

As an identity, Eq.~\eqref{idG} should be solved by any configuration of $\pi$. To further simplify our discussion, it is sufficient to choose a diagonal configuration for $\Pi^{\mu}_{\nu}$. (An alternative point of view, which also works for our purposes, is that $\Pi^{\mu}_{\nu}$ can always be diagonalized via an appropriate coordinate transformation around a given point.) Suppose $\{ \lambda_{0},\lambda_{1},\cdots,\lambda_{D-1} \}$ are the diagonal components of $\Pi^\mu_\nu$. Then, the ${}_{ {\bi\bi}}{}^{\ai\ai} (\ai\ne \bi, ~{\rm no~summation~for~\ai\ai~and~\bi\bi})$ component of $\widetilde{\G}_{\mu\nu}{}^{\rho\sigma}=0$ gives
\begin{align}
\widetilde{\G}^{N}_{ {\bi\bi}}{}^{ {\ai\ai}} =\G^{N}_{ {\bi\bi}}{}^{ {\ai\ai}} =	\eta^{ {\ai\ai}}{\pd\over\pd\lambda_\ai}G^{N}_{ {\bi\bi}} \equiv 0   .
\end{align}
Since $\eta^{\ai\ai}=\pm 1$, we have $\pd G^{N}_{ {\bi\bi}}/\pd \lambda_\ai=0$. Thus, $G^{N}_{ {\bi\bi}}$ must be independent of $\lambda_\ai(\ai\neq \bi)$. Since $G^{N}_{ {\bi\bi}}$ is $N$-th order in $\Pi$, it must be of the form
\begin{align}\label{eq:gnbb}
G^{N}_{\bi\bi}=C_\bi\lambda_\bi^N    ,
\end{align}
where $C_\bi$ is a constant. It follows from Lorentz invariance that the only possible form of $C_\bi$ should be $C\eta_{\bi\bi}$, with $C$ being a constant. Since we have $(\Pi^N)^{\mu}_{\nu} = \lambda^N_\mu \delta^\mu_\nu$ (no summation for $\mu$) for the diagonal configuration chosen, we must have
\begin{align}
\label{GNC}
G^{N}_{\mu\nu}=C\, \eta_{\mu\ri} (\Pi^N)^{\ri}_{\nu} =C\,(\Pi^N)_{\mu\nu}    .
\end{align}
 Again, by Lorentz invariance, these relations must be also satisfied by the non-diagonal components of  $\Pi^\mu_\nu$.

Therefore, there is only one term, $p_N g_{\mu\ri} (\K^N)^\ri_{\nu}$, at $N$-th order
that survives the consistency check in the decoupling limit, and we end up with
\begin{equation}
G_{\mu\nu} = \sum_N G^{N}_{\mu\nu}=g_{\mu\ri}\sum_N p_N(\K^N)^{\ri}_{\nu}    .
\label{eq:ansatzDL}   
\end{equation}
where $p_N$ now are constant. In summary, we have reduced our ansatz in Eq.~\eqref{eq:generalansatz} to
the one in Eq.~\eqref{eq:ansatzDL} by requiring that the non-minimal matter coupling does not give rise to Ostragradski ghosts in the decouling limit.

\subsubsection{Example: Lowest orders}\label{app:lowestorder}

Before moving on to the next step of the proof, it is instructive to give a concrete example to illustrate how these abstract arguments work in essence. Consider the most general composite metric, up to 2nd-order in $\K$:
\begin{align}
G_{\mu\nu}=&g_{\mu\nu}+a_1 \Tr{\K}g_{\mu\nu}+a_2 \K_{\mu\nu}+b_1\Tr{\K^2}g_{\mu\nu}	\nn
&+b_2 \Tr{\K}^2g_{\mu\nu} +b_3 \Tr{\K}\K_{\mu\nu}+b_4 \K_{\mu\rho}g^{\rho\sigma}\K_{\sigma\nu}    ,
\end{align}
where $a_i$ and $b_i$ are constants and $\K^{\mu}_{\nu}$ is to be evaluated in the decoupling limit as $\Pi^\mu_\nu$.  The identity in Eq.~\eqref{idG} can be straightforwardly calculated
\begin{align}
\widetilde{\G}_{\mu\nu}{}^{\rho\sigma}&= a_1 \eta^{\rho\sigma}\eta_{\mu\nu} -\frac{2a_1}{D+1}\delta^\ri_{(\mu} \delta^{\si}_{\nu)} - \frac{8b_1+4b_3}{D+2}\Pi^{(\ri}_{(\mu} \delta^{\si)}_{\nu)}
\nn
 &~~~ + 2b_1 \Pi^{\rho\sigma}\eta_{\mu\nu} + 2b_2\Tr{\Pi}\eta^{\rho\sigma}\eta_{\mu\nu}	+b_3 \eta^{\rho\sigma}\Pi_{\mu\nu} 
 \nn
 & ~~~+ \frac{4b_1-4b_2(D+2)+2b_3}{(D+1)(D+2)}\Tr{\Pi}\delta_{(\mu}^\rho\delta_{\nu)}^\sigma  
 \label{Gexp2nd}\nn
 &=0\ .
\end{align}
When $D>2$, all the terms in Eq.~\eqref{Gexp2nd} cannot cancel each other, so Eq.~\eqref{Gexp2nd} enforces $a_1=b_1=b_2=b_3=0$. Thus, up to 2nd order, the consistency requirement in the decoupling limit implies
\begin{equation}
G_{\mu\nu}=g_{\mu\nu}+a_2 \K_{\mu\nu}+b_4 \K_{\mu\rho}g^{\rho\sigma}\K_{\sigma\nu}\ .
\end{equation}
When $D=2$, we can get the same result, but one needs to take into account the Cayley--Hamilton theorem when checking cancellations between the terms in Eq.~\eqref{Gexp2nd}. (The Cayley--Hamilton theorem states that: suppose that $p(\lambda)=0$ is the characteristic polynomial of matrix $A$, then substituting $A$ for $\lambda$ in the polynomial gives rise to an identity, $p(A)=0$. To make use of this identity in Eq.~\eqref{Gexp2nd}, one can differentiate the identity with respect to A: $\pd p(A)/\pd A=0$.) When $N=D$, the Cayley--Hamilton identity is used directly; when $N>D$, one multiplies $p(A)=0$ with powers of $A$ to get relevant identities. The diagonalization of $\Pi^\mu_\nu$ in the last subsection, on the other hand, is a convenient way to avoid the complications due to the Cayley--Hamilton identities for $N\ge D$.

\subsection{No BD ghost from matter loop corrections}

Given an effective composite metric $G_{\mu\nu}$ in dRGT massive gravity or bigravity, it is natural to include the cosmological term $\sqrt{-G} \Lambda$, $\Lambda$ being constant, in the Lagrangian. If it is not there in the bare Lagrangian, it has been shown that matter loop corrections will generically generate a cosmological term for the effective metric~\cite{deRham:2013qqa}, much like that in general relativity. So, to avoid matter quantum corrections to re-introduce the BD ghost, we require
\be
\label{GUs}
\sqrt{-G}= \sqrt{-g} \sum_{s=0}^{D} a_s U_s(\mc{K})    ,
\ee
where $a_s$ are constants and $U_s(\mc{K})$ are defined in Eq.~\eqref{dRGTm}. We will show that this requirement  is sufficient to reduce
\begin{equation}
\label{ans2}
G_{\mu\nu} = \sum_N G^{N}_{\mu\nu}=g_{\mu\ri}\sum_N p_N(\K^N)^{\ri}_{\nu}    
\end{equation}
to
\be
\label{actbgeff}
G_{\mu\nu} = g_{\mu\ri} \big( \alpha^2 \delta^\ri_\nu+2 \alpha\beta\mc{K}^\ri_\nu + \beta^2(\mc{K}^2)^\ri_\nu \big)\ .
\ee

First, notice that $p_0$ should be non-zero (positive definite if the signature of the metric is taken into account), otherwise the effective metric $G_{\mu\nu}$ becomes singular in the limit $\mc{K}^\mu_\nu|_{g_{\mu\nu}\to f_{\mu\nu}}\to 0$. Therefore, we can re-write
\bal
\label{geffConst}
G_{\mu\nu}  &=  p_0 \,g_{\mu\ri} \left( \delta^\ri_\nu + P^\ri_\nu(\mc{K}) \right) 
\\
&= p_0\, g_{\mu\ri} \left( \delta^\ri_\nu + \sum_{N=1} p'_N \; (\mc{K}^N)^\ri_\nu \right)   ,
\eal
where $p'_N=p_0^{-1}p_N$. Note that $p_N$ are constant here. Then the determinant of ansatz (\ref{geffConst}) can be re-cast as
\bal
\sqrt{-G} &= p_0^{\frac{D}{2}} \sqrt{-g} \det( \sqrt{1+ P(\mc{K})} )
\\
&\equiv p_0^{\frac{D}{2}} \sqrt{-g} \det( {1+ Q(\mc{K})} )
\\
&=p_0^{\frac{D}{2}} \sqrt{-g} \left(\sum_{s=0}^D U_s( Q(\mc{K}) )\right)   ,
\eal
where
\be
Q^\mu_\nu(\mc{K}) = \sum_{N=1} q_N (\mc{K}^N )^\mu_\nu   ,
\ee
and $q_N$ can be expressed in terms of $p'_N$ by Taylor expanding $\sqrt{1+ P(\mc{K})}$ and comparing to the coefficients of $(\mc{K}^N )^\mu_\nu$. For the requirement (\ref{GUs}) to go through, the following equation
\be
\label{leftrightcomp}
\sum_{s=0}^D U_s( Q(\mc{K}) ) = \sum_{s=0}^D a_s U_s(\mc{K})
\ee
should be satisfied for some constant $a_s$. We will check what this requirement implies order by order in $\mc{K}$.

The $0$-th order equation can be satisfied by setting $a_0=1$. At order 1, we have $q_1 U_1(\mc{K}) = a_1 U_1(\mc{K})$, which gives $q_1=a_1$. At order 2, we have
\be
q_1^2 U_2(\mc{K}) + q_2 U_1(\mc{K}^2) = a_2 U_2(\mc{K})    ,
\ee
which leads to 
\be
q_1^2 = a_2, ~~~~q_2=0\ .
\ee
At order 3, making use of the fact that $q_2=0$, we have
\be
q_1^3 U_3(\mc{K}) + q_3 U_1(\mc{K}^3) = a_3 U_3(\mc{K})   ,
\ee
which leads to 
\be
q_1^3 = a_3, ~~~~q_3=0\ .
\ee
This can be extended to arbitrary orders, so that, at order $s$, we simply have
\bal
1<s\le D  &:  q_1^s U_s(\mc{K}) + q_s U_1(\mc{K}^s) = a_s U_s(\mc{K})   ,
\\
s> D  &:  q_s U_1(\mc{K}^s) = 0   .
\eal
That is, by solving Eq.~\eqref{leftrightcomp} order by order, we can conclude that
\be
q_1= a_1, ~~~ q_s=0 ~~~({s>1})\ ,
\ee
which leads to
\be
G_{\mu\nu} = p_0g_{\mu\ri} \big[ (1+q_1 \mc{K})^2 \big]^\ri_\nu\ .
\ee
Therefore, by requiring the BD ghost does not re-emerge under matter loop corrections, we have reduced the ansatz in Eq.~\eqref{ans2} to Eq.~\eqref{actbgeff}, as advertised. This is precisely the effective composite metric initially proposed in Ref.~\cite{deRham:2014naa}, and it emerged here naturally from requiring absence of {Ostrogradski ghosts in the decoupling limit and absence of the BD ghost from matter loop corrections}.

\section{Conclusion}
In this letter we have explored a generic class of composite couplings to matter in dRGT massive gravity and bigravity, which involve a generic admixture of the metrics $g_{\mu\nu}$ and $f_{\mu\nu}$. We have imposed two diagnostic tests to ensure the ghost-freeness of the theory at least below the strong coupling scale.
First, we have required such non-minimal coupling not to give rise to Ostrogradki ghosts for the scalar \stu \ mode in the decoupling limit. This has allowed us to discard a big subset of all such couplings.  Furthermore, we have imposed that matter loop corrections do not re-introduce the BD ghost, which has allowed us to single out one composite metric with two free parameters as the unique non-minimal coupling to matter. This is precisely the composite metric proposed in Ref.~\cite{deRham:2014naa}, and given in the metric language in Eq.~\eqref{actbgeff}. Consequently, cosmological solutions in these theories only depend on a finite choice of healthy couplings to matter at energy scales comparable to the strong coupling scale.

We note that our proof does not assume any specific form of the matter fields in the non-minimal matter coupling -- One can view $\psi$ as a vector field encompassing all possible fields. However, we do assume that matter fields are not coupled to derivatives of the metrics and in each matter sector all matter fields couple to one universal composite metric. More generically, however, one may consider a case where derivatives of the metrics also enter the matter couplings and matter fields couple to the two metrics in a convoluted non-trivial way. Whether ghost free non-minimal couplings generically exist in this case is beyond the scope of this letter, but such an exotic theory should exist and be free of the BD ghost below the strong coupling scale, based on the results of Ref.~\cite{Hinterbichler:2015yaa,deRham:2015cha}\,\footnote{We would like to thank Kurt Hinterbichler for discussions on this. 

}:  The non-minimal matter coupling (\ref{actbgeff}) has a very simple representation in the vielbein formulation. Taking this vielbein non-minimal matter coupling as the starting point, one can reproduce the non-minimal coupling in the metric formulation (\ref{actbgeff}), if one imposes the symmetric vielbein condition~\cite{deRham:2015cha}. On the other hand, if one imposes a modified vielbein condition, as in \cite{Hinterbichler:2015yaa}, one would end up with a convoluted metric theory that is physically different from the theory with  (\ref{actbgeff}). However, despite being complicated and exotic, the theory with the modified vielbein condition has the same decoupling limit as the theory with the symmetric vielbein condition~\cite{deRham:2015cha}.

\vskip 20pt

\noindent{\it Note added: The uniqueness of the composite metric in the vielbein formulation has been argued in Ref.~\cite{deRham:2015cha}, which appeared when our paper was being finalized. Our proof in the metric formulation is complementary to the comments in Ref.~\cite{deRham:2015cha}.}

\acknowledgements{We would like to thank Claudia de Rham, Kurt Hinterbichler, Rachel Rosen and Andrew J.~Tolley for useful discussions. RHR acknowledges the hospitality of the Perimeter Institute of Theoretical Physics during stages of this work. RHR's research was supported by a DOE grant de-sc0009946, the STFC grant ST/J001546/1 and in part by Perimeter Institute for Theoretical Physics. Research at Perimeter Institute is supported by the Government of Canada through Industry Canada and by the Province of Ontario through the Ministry of Economic Development \& Innovation. SYZ acknowledges support from DOE grant DE-SC0010600, and would like to thank Kavli Institute for Theoretical Physics China at the Chinese Academy of Sciences for hospitality during part of this work. QGH, YHX and KCZ are supported by the grants from NSFC (grant NO. 11322545 and 11335012). 

}
	
\appendix

\section{Projector $\mc{P}$}
\label{app:projop}

Consider a generic $(2,2)$ tensor $\Lambda^{\rho\sigma}_{\mu\nu}$ where the up and down two indices are symmetric respectively. We want to derive a projector that projects out the traces in $\Lambda^{\rho\sigma}_{\mu\nu}$. That is, for generic $\Lambda^{\rho\sigma}_{\mu\nu}$, we need $(\mc{P}\Lambda)^{\rho\nu}_{\sigma\nu}=0$.  Considering the index structure of $\Lambda^{\rho\sigma}_{\mu\nu}$, the only trace terms are $\Lambda^{\ai}_{\ai}{}^{(\rho}_{(\mu}\delta^{\sigma)}_{\nu)}$ and $\Lambda^{\ai}_{\ai}{}^{\bi}_{\bi} \delta^{(\rho}_{(\mu}\delta^{\sigma)}_{\nu)}$. Therefore we have
\begin{equation}\label{eq:plambda}
(\mc{P}\Lambda)^{\rho\sigma}_{\mu\nu}=\Lambda^{\rho\sigma}_{\mu\nu}+a\Lambda^{\ai}_{\ai}{}^{(\rho}_{(\mu}\delta^{\sigma)}_{\nu)}+b \Lambda^{\ai}_{\ai}{}^{\bi}_{\bi} \delta^{(\rho}_{(\mu}\delta^{\sigma)}_{\nu)}\ ,
\end{equation}
where $a$ and $b$ are constants. To determine $a$ and $b$, we impose the condition that the right hand side of Eq.~(\ref{eq:plambda}) vanish identically when $\sigma$ and $\nu$ are contracted. This leads to, in $D$ dimensions,
\begin{align*}
a=-\frac{4}{D+2}&, & b=\frac{2}{(D+1)(D+2)}\ .
\end{align*}

	\bibliographystyle{JHEPmodplain}
	\bibliography{references}

\end{document}